\documentclass[a4paper,11pt]{article}
\usepackage[utf8]{inputenc}
\usepackage{psfrag}
\usepackage{comment}
\usepackage{array}
\usepackage{amssymb}
\usepackage{amsmath}
\usepackage{amsthm}
\usepackage{graphicx}
\usepackage{caption}
\usepackage{hyperref}
\usepackage{float}
\usepackage{subfig}
\usepackage{authblk}
\usepackage[toc]{appendix}
\usepackage{color}

\usepackage{epsfig}

\numberwithin{equation}{section}

\title{Towards Timelike Singularity via AdS Dual}
\author{Samrat Bhowmick\thanks{email: tpsb5@iacs.res.in}}
\affil{\normalsize Department of Theoretical Physics, \authorcr Indian Association for the Cultivation of Science,
\authorcr Kolkata 700 032. India.}
\author{Soumyabrata Chatterjee\thanks{email: soumyac@iopb.res.in}}
\affil{\normalsize Institute of Physics, \authorcr Sachivalaya Marg, \authorcr Bhubaneswar  751 005. India.\authorcr and
\authorcr Homi Bhabha National Institute, \authorcr Training School Complex,Anushakti Nagar,Mumbai,India-400 085}

\begin{document}

\maketitle

\begin{abstract}
 \noindent It is well known that Kasner geometry with space-like singularity can be extended to bulk AdS-like geometry, furthermore one
can study field theory on this Kasner space via its gravity dual. In this paper, we show that there exists a Kasner-like
geometry with timelike singularity for which one can construct a dual gravity description. We then study various extremal surfaces
including space-like geodesics in the dual gravity description. Finally, we compute correlators of highly massive operators in the boundary field theory with a geodesic
approximation.
\end{abstract}

\section{Introduction}
    Time dependent geometries have played a profound role in the study of cosmology. One of the primary motivations of studying time-dependent cosmological space-times 
is to understand various aspects of the big bang singularity. Correlation functions computed within the framework of gauge theories living on such space-times 
can in principle furnish us with a natural set of tools to probe into the structure of these singularities. 
However formulation of gauge theories on such backgrounds is an analytical challenge which quite often seems to be insurmountable. 
One way of investigating field theory and cosmology on such backgrounds is to resort to the AdS/CFT correspondence {\cite{Malda 1997}} or the gauge/gravity duality 
which provide us with the holographic machinery for evaluating quantities relevant to strongly coupled gauge theories. 
For pure five dimensional $AdS$ space-time the holographic duality connects weakly coupled type II-B string theory on $AdS_5 \times S^5$ to a strongly coupled 
$\mathcal{N}=4$ Super Yang-Mills theory (SYM) living on the 4-dimensional conformal boundary of the $AdS$. Replacing the Poincare metric on the boundary with any 
Ricci flat time-dependent geometry (arising as solutions of type IIB SUGRA) invokes supersymmetry breaking deformations of the SYM side. These deformations can become more intractable 
for anisotropically expanding time-dependent backgrounds such as Kasner background. Kasner geometry typically has a a space-like singularity in the past, 
where some of the directions shrink to zero sizes. This gives rise to very strong gravitational fluctuations. 
In recent years several aspects of strongly coupled gauge theories living on such time-dependent
boundary of the $AdS$ has been studied with partial successes in constructing 
holographic models \cite{Hertog:2005hu} -- \cite{Mefford:2016res}.

    The anisotropic nature of the four dimensional Kasner space-time is captured by the three
Kasner exponents satisfying the Kasner conditions. This class of geometries arises as the solution of the vacuum Einstein equations in 4-dimensions
corresponding to the Bianchi type I universe. A non-vacuum
generalization of this class of geometries is possible in the presence of a perfect fluid stress tensor whereby the Kasner conditions are modified, 
making them further suitable for 
the cosmological context {\cite{Awad:2007fj,Koyama:2001rf,Chatterjee:2016bhj}}. A detailed study of holographic two-point correlation functions, 
in boundary gauge theories for both the vacuum {\cite{Banerjee:2015fua}} and the non-vacuum cases {\cite{Chatterjee:2016bhj}}, have been studied using 
the geodesic approximation for large mass dimensions of the boundary operators. In a recent work \cite{Shaghoulian:2016umj}, a space-dependent version
of the Kasner geometry has been  discussed wherein the space-like singularity of the Kasner geometry is replaced by a time-like singularity.
The space-dependent Kasner geometry can be used to probe into time-like BKL singularities. 
The four dimensional time-like Kasner geometry can be obtained as the IR limit of a deformation of a planar $AdS_4$ Black hole geometry\cite{Ren:2016xhb}. 
Interestingly a four dimensional time-like Kasner geometry can be embedded in a five-dimensional AdS space-time and the resultant configuration
can be obtained as a solution of $9+1$ dimensional type-IIB Supergravity in the presence of a nonzero self-dual five-form field strength 
and a constant/vanishing dilaton profile. 
In contrast with the space-like singularities relevant to cosmology, the time-like case may be well suited for the study of black hole interiors. 
An example of time-like singularity can be found within the structure of Reissner-Nordström (RN) black holes. 
However the RN geometry being a non-vacuum geometry, its extension into the bulk is rather difficult.
Also it is perhaps impossible to probe inside the black hole horizon holographically \cite{Ishibashi:2017jde}.
In any case the near-singularity behaviour of RN black hole geometry can be approximated by the non-vacuum version of the time-like cousin of the Kasner geometry.  
Hence a natural question to ask is whether a time-like singularity can be probed by using the AdS/CFT correspondence?
Therefore, in parallel with time-dependent $AdS$-Kasner backgrounds {{\cite{Banerjee:2015fua,Chatterjee:2016bhj}}, one can compute correlation 
functions in the case of the space-dependent $AdS$-Kasner space-time and use these correlators to probe into the time-like singularity at $r=0$.
This is the aim in this paper.    

 Unlike it's time-dependent counterpart, the space-dependent AdS-Kasner geometry is static. Correlation functions computed on such space-times should 
in general be smooth and free from divergences arising as a consequence of the presence of a singularity. 
Hence we also do not expect phenomenon like ``particle creation'' to occur on such geometry.
In this paper we find that two point functions are indeed smooth near the singularity and there is conspicuous absence of pole in the correlators
for both positive and negative values of the Kasner exponents. These features of the present computations in turn lends justification to the computations done
in the time-dependent case.

 We pursue the approach undertaken in {\cite{Engelhardt:2014mea,Banerjee:2015fua,Engelhardt:2015gta}}.
Using the geodesic approximations we compute the two-point correlators for largely massive operators.
We first consider a radial slicing of the boundary at a constant distance from the singularity and calculate equal-time correlators on the boundary.
In the limit of the radial coordinate($r$) going to zero, we obtain the equal time two-point correlator near the singularity. 
This method is suitable for cases with negative Kasner exponents. 
 we next consider a geodesic in the closest possible vicinity of the time-like singularity. 
Surprisingly, in this case, we find the correlator goes to a constant value as the geodesic approaches the singularity.
This method is suitable for positive Kasner exponents.
In the present set up, it is possible to compute extremal co-dimension 2 surfaces and allow them 
to approach towards the singularity. According to the prescription given in \cite{Ryu:2006ef},
the area of this extremal surface is related to entanglement entropy between two different regions of the boundary field theory \cite{Ryu:2006bv,Ryu:2006ef}.
In the present context, these extremal surfaces for different combinations of the Kasner exponents is 
another  way of probing the structure of the time-like singularity. 
In the case of pure AdS, nature of this co-dimension 2 surface is well studied, see for example \cite{Hubeny:2012ry}.
In the present case, where the geometry is locally AdS, the nature of extremal surface near the singularity is a matter of detailed investigation. 
The computation of the actual area near the singularity turns out to be difficult.
In this article we discuss the nature of the extremal surface in the space-dependent $AdS$-Kasner background.

The organization of the paper is as follows. In section \ref{sec:brane} we discuss the embedding of the four dimensional space-dependent Kasner 
space-time with a timelike singularity in five-dimensional locally AdS space-time. Here we show that this $AdS$-Kasner geometry emerges from a 
$D3$-brane configuration coupled to a four form gauge field with a purely 
space-dependent profile.
In sections \ref{cod4} and \ref{cod1} we present a study of the timelike singularity of the boundary Kasner space-time. 
In section \ref{cod4}, this singularity is probed by a closely approaching geodesic or a co-dimension-3 curve. Within a certain approximation, 
it enables us to extract the correlator of the
dual field theory out of the information encoded in the bulk geometry. In section \ref{cod1}, we give a description of  the co-dimension 2 surface and 
discuss its relevance in evaluating holographic entanglement entropy near the singularity. 
In section \ref{conc}, we conclude and discuss some further possibilities.

\section{D3 brane with timelike Kasner world volume} \label{sec:brane}
Black branes geometries are well known in string theory and supergravity.
Besides the static D-branes of odd space dimensions, the IIB string theory admits {\it time dependent}
branes as discussed in \cite{Banerjee:2013jn}. Here world volume of the branes is Kasner-like. In this note, we present 
another solution of type IIB supergravity where 
 brane world volume is Kasner-like but with timelike singularity. 
Consider, for example, the case of 
D3 brane. The equations of motion following from the relevant part of standard IIB 
supergravity action
are of the form \footnote{The self-duality condition of the $5$-form field strength
is to be imposed at the level of the equation of motion.}
\begin{eqnarray}
&&R^{\mu}_{\nu} = \frac{1}{2} \partial^\mu \phi \partial_{\nu} \phi + \frac{1}{2 \times 5!} \left(5 F^{\mu \xi_2...\xi_5}
 F_{\nu \xi_2...\xi_5}- \frac{1}{2} \delta^\mu_\nu F_5^2\right),\nonumber\\
&&\partial_\mu({\sqrt{g}} F^{\mu \xi_2...\xi_5}) =0, \nonumber\\
&&\nabla^2\phi =0.
\label{2b}
\end{eqnarray}
These equations are solved by
\begin{eqnarray}
\label{tlK}
&&ds^2 = \Big(1 + \frac{l^4}{\rho^4}\Big)^{-\frac{1}{2}} \Big[ - r^{2p_t}dt^2 + r^{2p_1} dx_1^2
+ r^{2p_2} dx_2^2 +  dr^2\Big] + \Big(1 + \frac{l^4}{\rho^4}\Big)^{\frac{1}{2}}
\Big[d\rho^2 + \rho^2 d\Omega_5^2\Big], \nonumber \\
\\
\label{tlKF}
&&F_{tx_1x_2r\rho} = \frac{2\sqrt{2} l^4r^{p_t+p_1+p_2}\rho^3}{(l^4 + \rho^4)^2},
~~~~~
 F_{ijklm} = \sqrt{-g} \,\epsilon_{tx_1x_2r\rho ijklm}\, F^{tx_1x_2r\rho}
\\
\label{tlKphi}
&&\phi =0,
\label{timev}
\end{eqnarray}
provided 
\begin{equation}
p_t+p_1 + p_2 = 1~~~{\rm and}~p_t^2+p_1^2 + p_2^2 =1.
\label{restric}
\end{equation}
Here, $i, j, k, l, m$ are the indices on $S^5$.
Note here, the world volume has a non-trivial geometry $ds^2 =- r^{p_t}dt^2 + r^{2p_1} dx_1^2
+ r^{2p_2} dx_2^2 +  dr^2$. This geometry is Ricci flat but has non trivial singularity structure.
So full timelike Kasner-brane geometry \ref{tlK} has a timelike singularity at $r=0$, one can 
verify this by calculating Kretschmann scalar $R^{\mu\nu\lambda\sigma}R_{\mu\nu\lambda\sigma}$.
This singularity is a timelike singularity.

We now take {\em near horizon} limit, $\rho \to 0$ of above geometry.
We get
\begin{equation}
 \label{NewKADS1}
 ds^2 = \frac{\rho^2}{l^2} \Big[ - r^{p_t}dt^2 + r^{2p_1} dx_1^2
+ r^{2p_2} dx_2^2 +  dr^2\Big] + \frac{l^2}{\rho^2}d\rho^2 + l^2 d\Omega_5^2\Big]\;,
\end{equation}
with 
\begin{equation}
F_{tx_1x_2r \rho} = \frac{4 r \rho^3}{l^4}, ~~{\rm giving ~potential}~C_{tx_1x_2r} = \frac{ r \rho^4}{l^4}.
\end{equation}
The space-time geometry expressed in \ref{NewKADS1} is very similar to Kasner-$AdS_5 \times S^5$ found in \cite{Banerjee:2013jn}. 
We make a coordinate transformation as $z=\frac{1}{\rho}$. We take $l=1$ for convenience. In this 
coordinates our new Kasner-$AdS_5$ geometry becomes
\begin{equation}
 \label{NewKADS}
 ds^2 = \frac{1}{z^2} \left[ - r^{2p_t}dt^2 + r^{2p_1} dx_1^2
+ r^{2p_2} dx_2^2 +  dr^2 + dz^2\right]\;.
\end{equation}
The boundary of this geometry is at $z=0$
and it has a singularity at $r=0$, again one can see this from the expression of the Kretschmann scalar,
$R_{\mu\nu\lambda\sigma}R^{\mu\nu\lambda\sigma}$. 
The boundary of this geometry itself has this timelike singularity. We are going to 
investigate this singularity in this paper. We consider a quantum field theory resides
on the boundary of (\ref{NewKADS}). We may think of this QFT to be some deformed CFT
whose dual is given by (\ref{NewKADS}).

\section{Investigating Timelike Singularity}
Let us consider co-dimension 2 and co-dimension 4 extremal space-like surfaces in the boundary. 
Both of them extend into the bulk. The co-dimension 4 extremal space-like curve i.e. the geodesic, gives correlator
of an operator of the boundary conformal field theory
in the limit of large conformal dimensions, whereas extremal co-dimension 2 surface gives 
entanglement entropy in the same CFT.

\subsection{Extremal Space-like Co-dimension 4 Curve in bulk} \label{cod4}
We now consider a dual gauge theory in the boundary of the bulk geometry mentioned in the previous section.
We would like to compute $\langle \psi|\mathcal{O}(x'_1,r_0)\mathcal{O}(x''_1,r_0) |\psi \rangle$, where $|\psi\rangle$
and $\mathcal{O}(x,r_0)$ are a state and an operator of the strongly coupled theory residing on the boundary.
We insert two operators in the $x_1$-direction at the points $x'_1$ and $x''_1$. This gives us an equal time correlator
at a fixed $r$. 

Here we consider only operator with large conformal dimensions, $\Delta = \frac{d}{2} + \sqrt{\frac{d^2}{4}+m^2}$, $d$ being
space-time dimensions, and large $N$ theory.
In this limit of the theory and the operators, a well known approximate
formula exists for the correlator \cite{Bala 1999} -- \cite{Ban 1998}. This formula in our case reads
\begin{equation}
 \langle \psi|\mathcal{O}(x',r_0)\mathcal{O}(x'',r_0) |\psi \rangle = e^{-m \mathcal{L}_{reg}(x',x'')} \;,
\end{equation}
where $\mathcal{L}_{reg}(x',x'')$ is the regularized length of the geodesic connecting $x'$ and $x''$. 
We already mentioned we would like to fix our two points in $x_1$ direction. That fixes time also and so 
we get equal time correlator. We choose two boundary points $x'_1$ and $x''_1$ at a fixed radial distance $r=r_0$.
Corresponding geodesic must then have two fixed end points $x'_1,x''_1$ at the boundary $z = 0$ at $r=r_0$.
For this particular calculation, therefore, the other boundary directions  $x_i, i \ne 1$ and $t$ are irrelevant. 
For the moment, we work with a general scale factor $a_1(r)$ along $x_1$. Later, we will
use the explicit form $a_1 = r^{p}$.

Here we follow a similar approach used in \cite{Banerjee:2015fua}.  
Now calling $x_1$ as $x$ and $a_1$ as $a$ for notational simplicity, the geodesic equations for (\ref{NewKADS})
are given by
\begin{eqnarray}
&&x^{\prime\prime} + 2 \frac{a^\prime}{a} x^\prime + a a^\prime {x^\prime}^3 = 0,\nonumber\\
&&z z^{\prime\prime} + {z^\prime}^2 + {x^\prime}^2 a^2 + a a^\prime z z^\prime {x^\prime}^2 +1 = 0.
\label{suthree}   
\end{eqnarray}
Here, we have taken radial coordinate, $r$ as a parameter and  $'$ denotes derivative are with respect to $r$.
General solutions of these equations can be written as
\begin{equation}
x (r) = \pm \int \frac{a(r_*) dr}{a(r) {\sqrt{a^2(r) -a^2(r_*)}}}
\label{sufour}
\end{equation}
and
\begin{equation}
z = + {\sqrt{ -2 \int dr  \left[\frac{a(r)}{\sqrt{a^2(r) - a^2(r_*)}} \left(\int^{r} dr' \frac{a(r')}{\sqrt{a^2(r') - 
a^2(r^{\prime}_*)}}\right)\right]
}}.
\label{sufive}
\end{equation}
In the above expressions $r_*$ is the turning point inside the bulk where we impose the boundary conditions that
$\frac{dr}{dx}\mid_{r=r_*}=0$ and $\frac{dz}{dx}\mid_{r=r_*}=0$.

One can see immediately that, with $a(r)=r^{2p}$, reality of above expression implies $0<r_*<r_0$ for $p>0$ and 
$0<r_0<r_*$ for $p<0$. So for $p>0$ we can vary $r_*$, the turning point, near singularity by keeping $r_0$(i.e the boundary
$z=0$)fixed, and for $p<0$ we can push $r_0$, the boundary, near singularity by keeping $r_*$(the turning point) fixed.
In this subsection we take $p$ to be negative. 
In the next subsection, we will study correlator for positive $p$.

\subsubsection{$p<0$ case} \label{cod4.1}

In case of $p<0$, our method is a bit different from \cite{Banerjee:2015fua}. The differential equations (\ref{suthree}) can be 
converted to hypergeometric differential equations with an argument $r^{2p}$. But here $p$ being negative and $r$ being close to $0$,
we need to solve this hypergeometric differential equation around $r=\infty$. That is one of Kummer's 24 solutions. Plugging
$p=-q$ in the differential equation (\ref{suthree}), we proceed to solve it and finally retrieve the solution by replacing $q=-p$. 
In appendix \ref{geosol}, we present a layout of this calculation.
The two integration constants appearing in the integration of the $x$-equation are fixed by the two boundary conditions namely a coordinate shift along the $x$-direction and 
by demanding $\frac{dr}{dx}\mid_{r=r_*}=0$.
For the $z$-equation, the corresponding boundary conditions are (1) $dz/dx\mid_{r=r_*}=0$ at the turning point of the geodesic in the $z-x$
plane and (2) requiring $z =0$ for $r = r_0$.

So with $a=r^p$, the integration in (\ref{sufour}) and (\ref{sufive}) can be performed. The answers turns out to be
\begin{equation}
 \label{x-noscale}
 x =\pm \frac{r^{1-p}}{1-p} \sqrt{\left(\frac{r}{r_*}\right)^{2p}-1} \;\left[1-{}_2F_1\left\{1,\frac{p-1}{2p},\frac{2p-1}{2p},\left(\frac{r}{r_*}\right)^{-2p}\right\}\right]
\;,
\end{equation}
and
\begin{eqnarray}
 \label{z-noscale}
 z &=& \left\{{\frac{2i\sqrt{\pi}{r_*}^{1+p}\Gamma(1-\frac{1}{2p})}{\Gamma(\frac{-1+p}{2p})}} r  \, _2F_1\left[\frac{1}{2},-\frac{1}{2 p},1-\frac{1}{2 p},\left(\frac{r}{r_*}\right)^{-2 p}\right] \right. \nonumber \\
        &-& \left. r^2 \, _3F_2\left[\left\{1,\frac{1}{2}-\frac{1}{2 p},-\frac{1}{p}\right\};\left\{1-\frac{1}{p},1-\frac{1}{2 p}\right\}
        ;\left(\frac{r}{r_*}\right)^{-2 p}\right] + c \right\}^{1/2} \;,
\end{eqnarray}
where the constant $c$ is determined by the condition $z=0$ at $r=r_0$. The integration constants have to be chosen in such a 
way that the full expressions of the right hand side of equations (\ref{x-noscale}) and (\ref{z-noscale}) become real.

We use now the scaling symmetry present in the 
geometry of (\ref{NewKADS}). The scaling symmetry of (\ref{NewKADS}) is 
\begin{equation}
 \label{ssymm}
 z \longrightarrow \lambda z, \, x_i \longrightarrow \lambda^{1-p_i} x_i, \, t \longrightarrow \lambda^{1-p_i} t, \,
 r \longrightarrow \lambda r \;.
\end{equation}
We use (\ref{ssymm}) with $\lambda = r_*$, to define new variable $(\bar z, \bar x, \bar r)$ as
\begin{equation}
 z = r_* \bar z,~~r = r_* \bar r, ~~x = {r_*}^{1-p} \bar x \;.
\end{equation}
In terms of these re-scaled variables, we can write
\begin{equation}
 \label{x-scale}
 \bar x =\pm \frac{r^{1-p}}{1-p} \sqrt{{r}^{2p}-1} \;\left[1-{}_2F_1\left\{1,\frac{p-1}{2p},\frac{2p-1}{2p},{\bar r}^{-2p}\right\}\right]
\;,
\end{equation}
and
\begin{eqnarray}
 \label{z-scale}
 \bar z &=& \left\{{\frac{2i\sqrt{\pi}\Gamma(1-\frac{1}{2p})}{\Gamma(\frac{-1+p}{2p})}} \bar r  
 \, _2F_1\left[\frac{1}{2},-\frac{1}{2 p},1-\frac{1}{2 p},{\bar r}^{-2 p}\right] \right. \nonumber \\
        &-& \left. {\bar r}^2 \, _3F_2\left[\left\{1,\frac{1}{2}-\frac{1}{2 p},-\frac{1}{p}\right\};\left\{1-\frac{1}{p},1-\frac{1}{2 p}\right\}
        ;{\bar r}^{-2 p}\right] + c \right\}^{1/2} \;.
\end{eqnarray}
Now $\bar r$ varies form $\bar r_0$ to 1.
These expressions undergo considerable simplification for some specific values of $p$. We choose such values to do
a further calculation.  

We intend to keep the discussion general to write down formal expressions. We would like to calculate 
the regularized geodesic length first. This is given by
\begin{equation}
 \label{geolen}
 \mathcal{L} = \int_{\bar r =1}^{\bar r = \bar r_0 - \delta }
 \frac{2 d \bar r}{\bar z(\bar r)} \sqrt{1+\left(\frac{d \bar z}{d \bar r}\right)^2+\bar r ^{2p}\left(\frac{d \bar x}{d \bar r}\right)^{2}}\;.
\end{equation}
This length turns out to be infinite for $\delta=0$. To regularize this we subtract from $\mathcal{L}$ the infinite
part of geodesic length of pure AdS. This removes $\delta \to 0$ singularity. Consequently
\begin{equation}
 \mathcal{L}_{\text{reg}} = \lim_{\delta \to 0} \left[\mathcal{L} - 2 \log{\frac{1}{\bar z(\bar r -\delta)}}\right]
\end{equation}
becomes finite and function of $r_0$ only. The correlator of our probe operator $\mathcal{O}$ becomes function of $r_0$ 
only because of the scaling symmetry,
\begin{equation}
 \langle \mathcal{O}(-x,r_0) \mathcal{O}(x,r_0) \rangle = \langle \mathcal{O}(-\bar x,\bar r_0) \mathcal{O}(\bar x,\bar r_0) \rangle 
 = e^{-m \mathcal{L}_{\text{reg}}} = f(\bar r_0) \;.
\end{equation}
 One can investigate now the nature of this correlator in the limit of
timelike singularity $r_0 \to 0$.

This general calculation can be done numerically, however we already mentioned before that the expressions
of $\bar x$ and $\bar z$ get simplified for specific values of $p$. So we have done explicit 
calculation of the regularized length for $p=-\frac{1}{4}$ and $p=-\frac{1}{6}$.  For $p=-\frac{1}{4}$,
$\bar x$ and $\bar z$ get simplified to
\begin{eqnarray}
 \label{x1by4}
 \bar x(\bar r) &=& \pm \frac{4}{15} \sqrt{1-\sqrt{\bar r}} \left(3 \bar r+4 \sqrt{\bar r}+8\right) \\
 \label{z1by4}
 \bar z(\bar r) &=& \frac{4}{3} \sqrt{\bar r^{3/2}-{\bar r}_0^{3/2}+3 \left(\bar r-\bar r_0\right)}.
\end{eqnarray}
Putting them back in (\ref{geolen}) one finds
\begin{equation}
 \mathcal{L} = 2 \tanh ^{-1}\left(\frac{\sqrt{1-\sqrt{\bar r_0-\delta }} \left(\sqrt{\bar r_0-\delta }+2\right)}{\sqrt{1-\sqrt{\bar r_0}}
   \left(\sqrt{\bar r_0}+2\right)}\right)\;,
\end{equation}
and $\mathcal{L}_{\text{reg}}$ turns out to be
\begin{equation}
 \mathcal{L}_{\text{reg}} = \log \left[\frac{64}{9} \left(-r_0^{3/2}-3 r_0+4\right)\right] \;.
\end{equation}
So the correlator
\begin{equation}
 \label{cor1}
 f(\bar r_0) = \left(\frac{64}{9}\right)^{-m} \left(-\bar r_0^{3/2}-3 \bar r_0+4\right){}^{-m} \;.
\end{equation}
Therefore as we go towards singularity, $\bar r_0 \to 0$, $f(\bar r_0)$ goes to $\left(\frac{256}{9}\right)^{-m}$,
but we consider here only highly massive operator. So we see correlator goes to 0 near singularity.
One should observe here that, at $\bar r_0 =1$, $f(\bar r_0)$ diverges, but it is not a problem because
it is an infrared divergence from gravity point of view. Similar behaviour can be observed for $p>0$ case for space-like
singularity as in \cite{Banerjee:2015fua}.
The reason for the occurrence of IR divergence is similar here. This particular singularity appears when $r_0$ goes close to the turning point. In our parametrization, the boundary ($z=0$) is situated at $r=r_0$.
In that case, the boundary is close to the turning point, 
and so the length contribution comes from the cutoff $\delta$. In $\delta\rightarrow 0$ limit the geodesic length approaches to zero. Normally this
produces a divergence in the two-point function. Since this divergence occurs from turning point, which is now very close to the boundary, we can interpret it as usual short distance divergence from dual field theory point of view.

We show another exponent $p=-\frac{1}{6}$ to illustrate the result more clearly. Again $\bar x$ and $\bar z$ get simplified.
One gets, after doing the integration (\ref{geolen}), 
\begin{equation}
 \mathcal{L} = 2 \tanh ^{-1}\left(\frac{\sqrt{1-(\bar r_0-\delta )^{1/3}} \left\{3
   \left(\bar r_0-\delta \right){}^{2/3}+4 \left(\bar r_0-\delta \right){}^{1/3}
   +8\right\}}{\sqrt{1-{\bar r_0}^{1/3}} \left(3 {\bar r_0}^{2/3}+4
   {\bar r_0}^{1/3}+8\right)}\right) \;,
\end{equation}
And consequently regulated length,
\begin{equation}
\mathcal{L}_{\text{reg}} = \log \left[\frac{16}{25} \left(64-9 \bar r_0^{5/3}-15 \bar r_0^{4/3}-40
   \bar r_0\right)\right] \;.
\end{equation}
So the correlator, $f(\bar r_0)$ turns out to be
\begin{equation}
  f(\bar r_0) = \left(\frac{16}{25}\right)^{-m}\left(64-9 \bar r_0^{5/3}-15 \bar r_0^{4/3}-40 \bar r_0\right){}^{-m} \;.
\end{equation}
In the limit $\bar r_0 \to 0$ this becomes $\left(\frac{1024}{25}\right)^{-m}$, which is almost 0 for highly
massive operators. We give here a plot of correlator. We choose $m=5$ for illustration purpose, for high value
of mass it will go faster.
\begin{figure}[H]
 \centering
 \includegraphics[width=.7\textwidth]{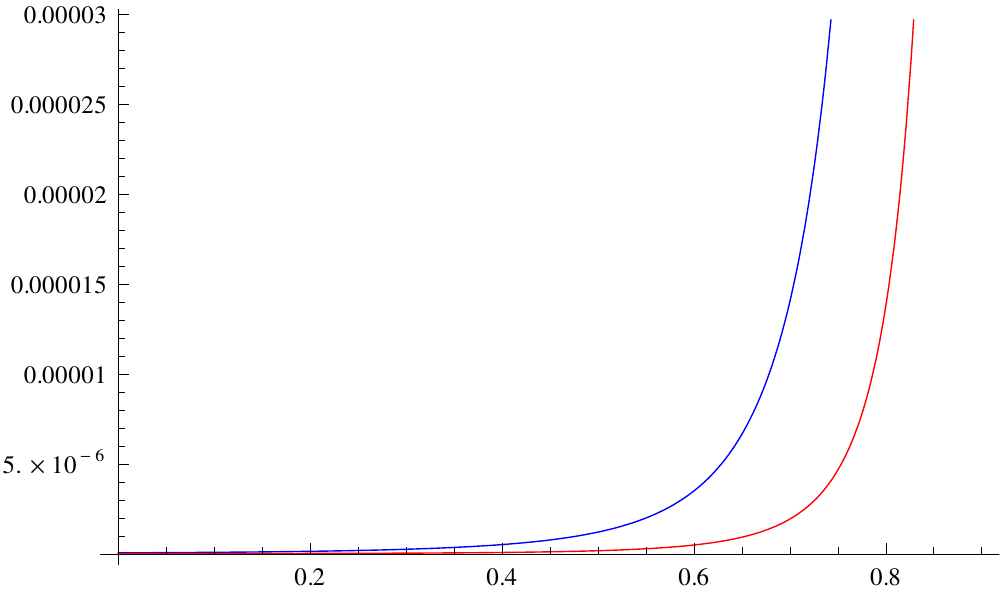}
 \caption{They are plots of $f(\bar r_0)$ vs. $\bar r_0$. Blue and red graphs are for $p=-\frac{1}{4}$
 and $p=-\frac{1}{6}$ respectively, $m=5$.}
  \label{fig:cor1}
\end{figure}

\subsubsection{$p>0$ case} \label{cod4.2}

Now we consider the boundary direction whose corresponding Kasner exponent is positive and therefore signify an expansion along that direction.
In order to investigate the behaviour of correlators near the singularity we follow the prescriptions of
\cite{Engelhardt:2014mea,Engelhardt:2015gta}. For the sake of brevity we fix the boundary at the radial slice $r= r_0 =1$,
and the turning point of the probe geodesic in the bulk at $r=r_*$ where we assume $r_* \to 0$. For definiteness, we take $p=\frac{1}{3}$.
The geodesic equations read
\begin{eqnarray}
 \label{pvegeoX}
 r x'' + \frac{2}{3} x' + \frac{1}{3} r^{2/3} x'^3=0 \;, \\
 \label{pvegeoZ}
  z z'' + 1 +z'^2+r^{2/3} x'^2 + \frac{1}{3} r^{-1/3} z z' x'^2 = 0 \;.
\end{eqnarray}
Here the boundary $z=0$ is at $r=1$. The boundary conditions $\frac{dz}{dx}|_{r=r_*}=0$, $\frac{dr}{dx}|_{r=r_*}=0$
and $x(r_*) =0$ fix $x$ and $z$ uniquely as
\begin{eqnarray}
  x(r)=\pm {3{r_*}^{2/3}\sqrt{\left(\frac{r}{r_*}\right)^{2/3}-1}} \;, \\
  z(r)=\sqrt{\left(1-r^2\right)+3 {r_*}^{2/3}\left(1-r^{4/3}\right)} \;.
\end{eqnarray}
We can now compute the geodesic length between two points, $-x$ to $x$ on the boundary. This is given by
\begin{equation}
\mathcal{L}= 2 \int_{r=r_*}^{1-\delta}\frac{1}{z(r)}\sqrt{1+r^\frac{2}{3}\left(\frac{dx}{dr}\right)^2+\left(\frac{dz}{dr}\right)^2}\,dr \;.
\end{equation}
Note that here we set a cut off near boundary at $r=1-\delta$.

To regularize the length of the geodesic we subtract  from the divergent piece appearing in the unregulated length, the equivalent AdS part and
arrive at the following result, namely,
\begin{equation}
 \mathcal{L}_{\text{reg}}=\lim_{\delta \to 0} \left[ \mathcal{L}-2 \log \left(\frac{1}{z(1-\delta)}\right)  \right]
 = \log \left(4+12 r_*^{2/3}-16 r_*^2\right) \;.
\end{equation}
The boundary correlator can now be written in terms of the regularized length of the geodesic as, 
\begin{equation}
 \label{cor2}
 \langle \mathcal{O}(-x, r_*) \mathcal{O}( x, r_*) \rangle = \left(4+12 r_*^{2/3}-16 r_*^2\right)^{-m} \;.
\end{equation}
The correlator has a usual short distance singularity for $r_*=1$ but when we probe the gravitational singularity by taking $r_*$
towards 0, we find no singular behaviour. It goes to $4^{-m}$, which is $0$ for large mass dimensions of the boundary operator. In figure (\ref{fig:cor2})
we demonstrate the spatial 
behaviour of the correlator through a plot.

\begin{figure}[H]
 \centering
 \includegraphics[width=.7\textwidth]{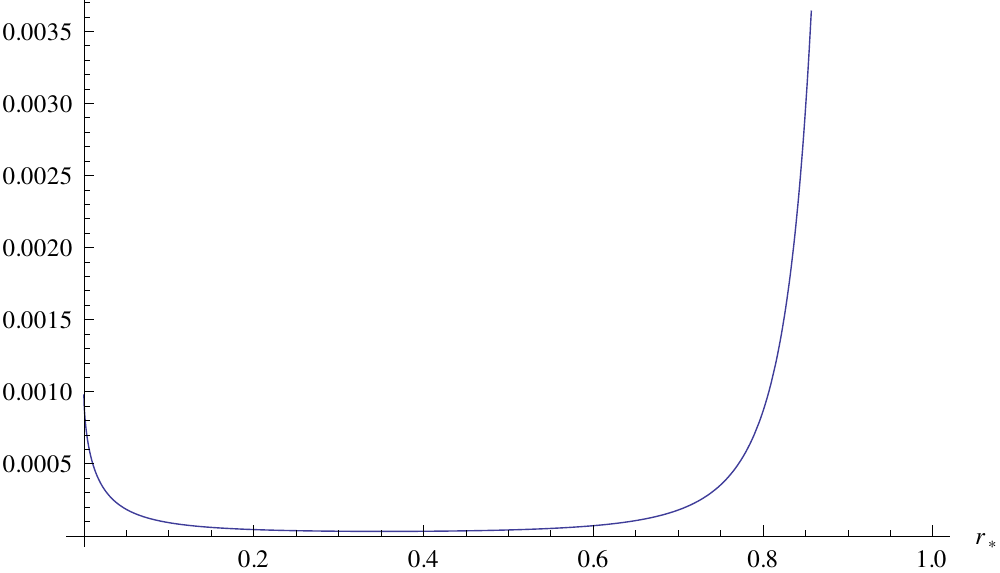}
 \caption{Here we plot $\langle \mathcal{O}(-x, r_*) \mathcal{O}( x, r_*) \rangle $ vs. $r_*$ for $p=\frac{1}{3}$ with $m=5$.}
  \label{fig:cor2}
\end{figure}

\subsection{Extremal Space-like Co-dimension 2 Surface in bulk} \label{cod1}
In this section, we discuss properties of space-like surfaces near the singularity. These surfaces are important 
in the context of entanglement entropy via a AdS dual. In \cite{Netta 2013} the authors computed entanglement entropy for
a confining gauge theory living on a time dependent Kasner space-time.
 Here we consider a space-like surface in the background of (\ref{NewKADS}). We consider the following region in the boundary: 
 $x_1 \in [-\infty,\infty]$, $x_2 \in [-\infty,\infty]$ and $r \in [r_0-\Delta r_0,r_0+\Delta r_0]$.
This surface may extend into the $z$-direction. To find the area we need to find the induced metric 
$G_{ab}=g_{\mu\nu}\frac{\partial X^\mu}{\partial \xi^a}\frac{\partial X^\nu}{\partial \xi^b}$ on this surface,
where $\xi^a$ and $\xi^b$ are coordinates on the surface. The area is given by
\begin{equation}
 A = \int d^3\xi \sqrt{G} \;.
\end{equation}
We choose a gauge where $\xi^1 = x_1, \xi^2 = x_2$ and $\xi^3 = r$.
In this gauge area functional turns out to be
\begin{equation}
 \label{AF}
 A = V \int dr \frac{r^{p_1+p_2} \sqrt{1+z'(r)^2}}{z(r)^3} \,,
\end{equation}
where $V$ is the volume of $x_1$ and $x_2$ directions. To find extremal surface one
has to find Euler-Lagrange equation for area functional (\ref{AF}). It turns out to be
\begin{equation}
 \label{EL}
 r z z''+\left[1+z'^2\right] \left[\left(p_1+p_2\right) z z'+3 r\right] = 0 \;.
\end{equation}
In the case of pure AdS, $p_1=p_2=0$ and so the Euler-Lagrange equation reduces to a much simpler form which can 
be solved exactly \cite{Hubeny:2012ry}.
In our case, it is difficult to solve the equation of motion analytically. Hence we resort to numerical means. In order to 
fix the boundary conditions we make use of the fact that the cross-section of the surface in the $z-r$ plane 
has a turning point at the point $z=z_*$ where
$\frac{dz}{dx}|_{z=z^*} = 0$. We initially fix the boundary at a radial slice $r=r_0$ which is finally pushed towards the singularity $r=0$.  
\begin{figure}[H]
 \centering
 \includegraphics[width=.5\textwidth]{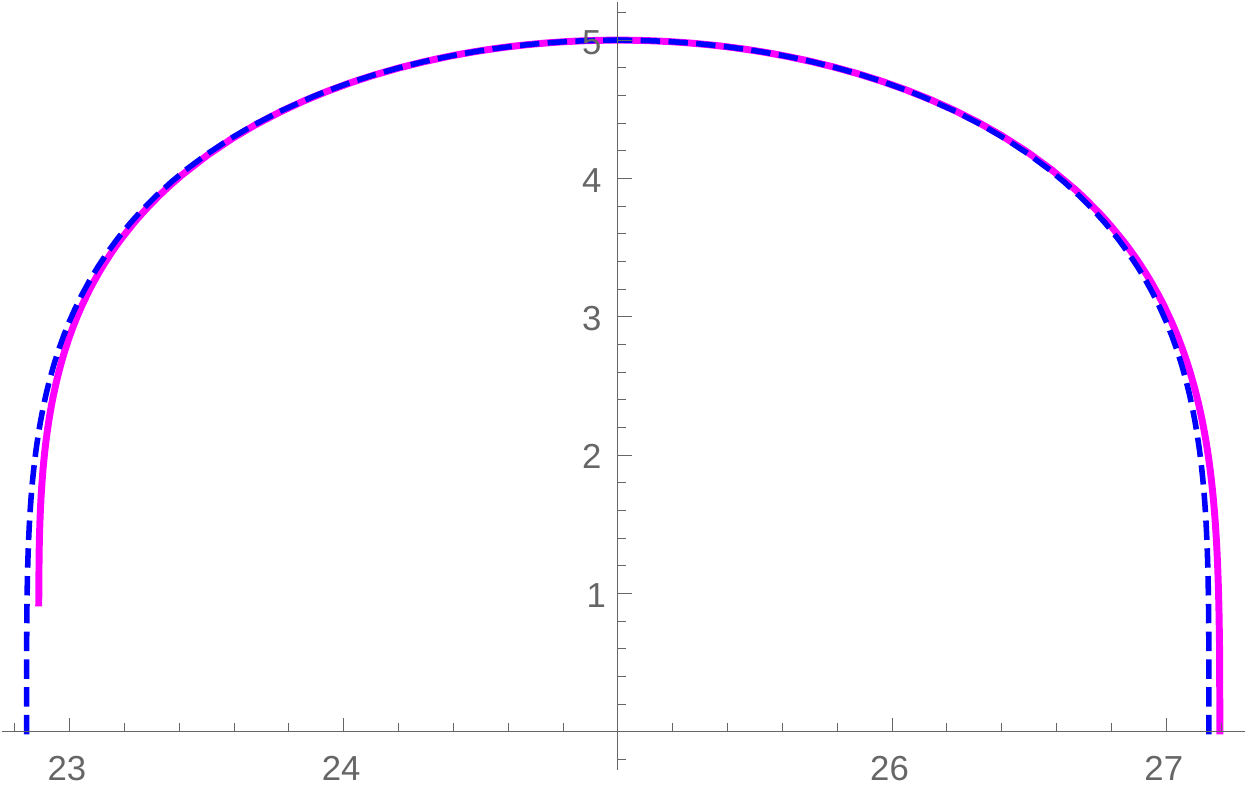}
 \caption{Here we plot the cross-section of the extremal surface in $z-r$ plain. The magenta curve is the surface described
 by \ref{EL} whereas dashed blue curve shows that of a pure AdS. In this plot, $z_*=5$ and $r_0=25$.}
 \label{fig:1}
\end{figure}
Note that the surface described by \ref{EL} is slightly shifted from the extremal surface
of pure AdS. It is actually expected, because of the presence of $r^{p_1+p_2}$ in the area functional (\ref{AF}).
In the figure above we take $p_1+p_2=0.33$. Note that as $p_1+p_2$ becomes smaller, the two curves tend to coincide
with each other. 

In Figure \ref{fig:1}, we take $r_0=25$. It is clearly seen from the graphs below that, as we decrease $r_0$, the 
shift becomes more prominent. But the nature of the curves remain almost same as we go towards the singularity. 
\begin{figure}[H]
 \centering
 \subfloat[]{
 \includegraphics[width=.3\textwidth]{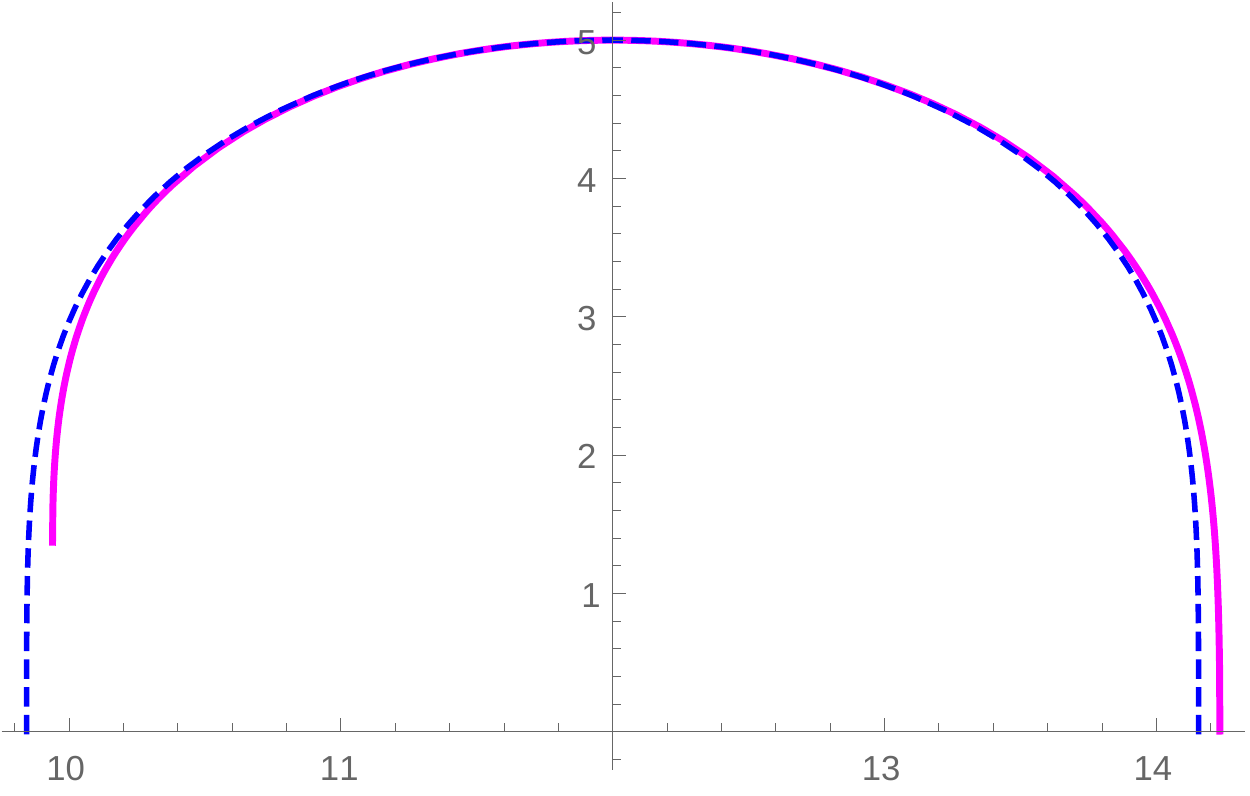}}
 ~~~~~~~~~~~
 \subfloat[]{
 \includegraphics[width=.3\textwidth]{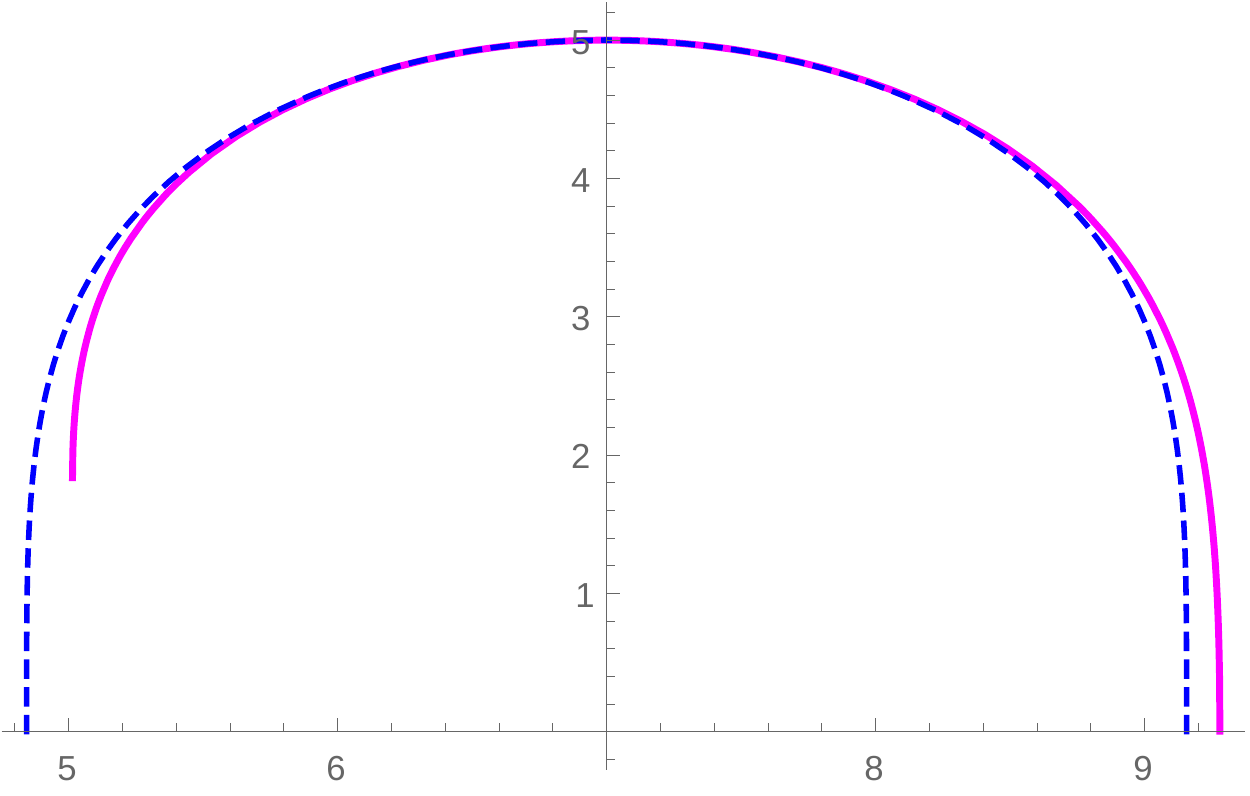}}
 
 \subfloat[]{
 \includegraphics[width=.3\textwidth]{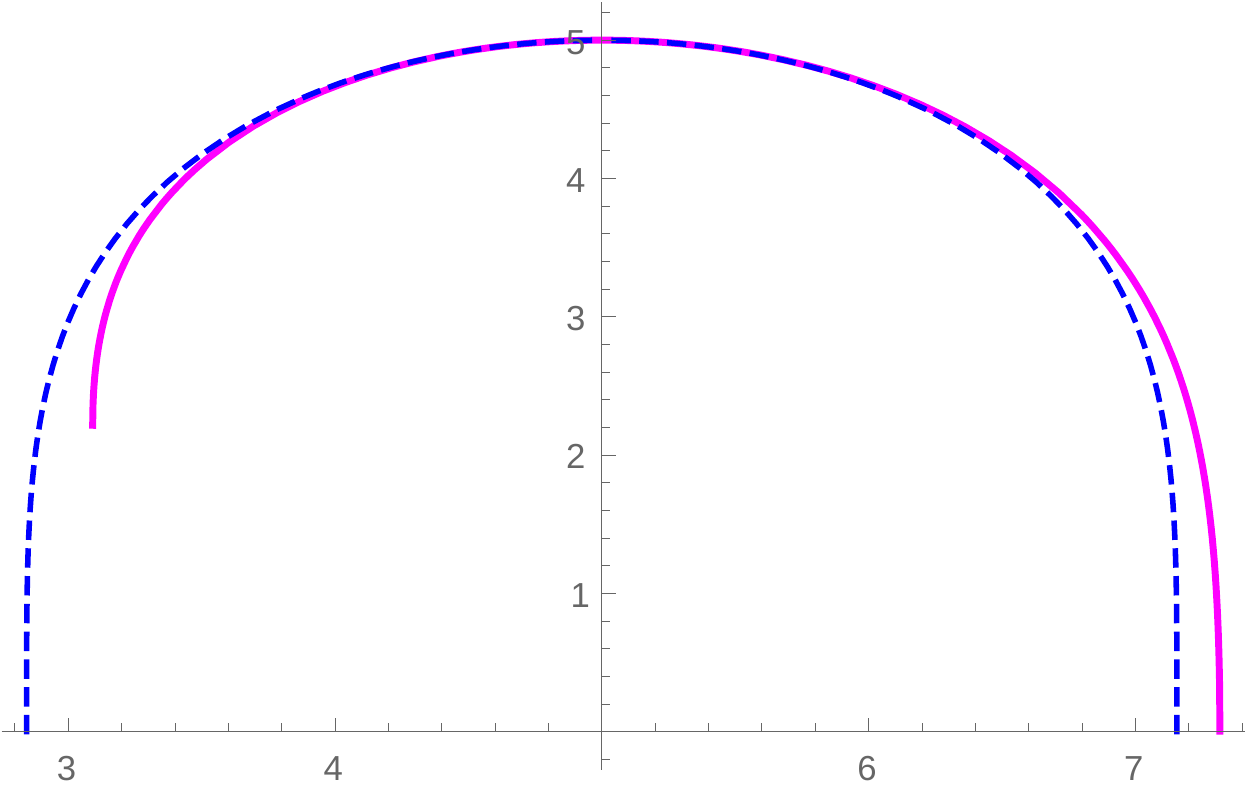}}
 ~~~~~~~~~~~
 \subfloat[]{
 \includegraphics[width=.3\textwidth]{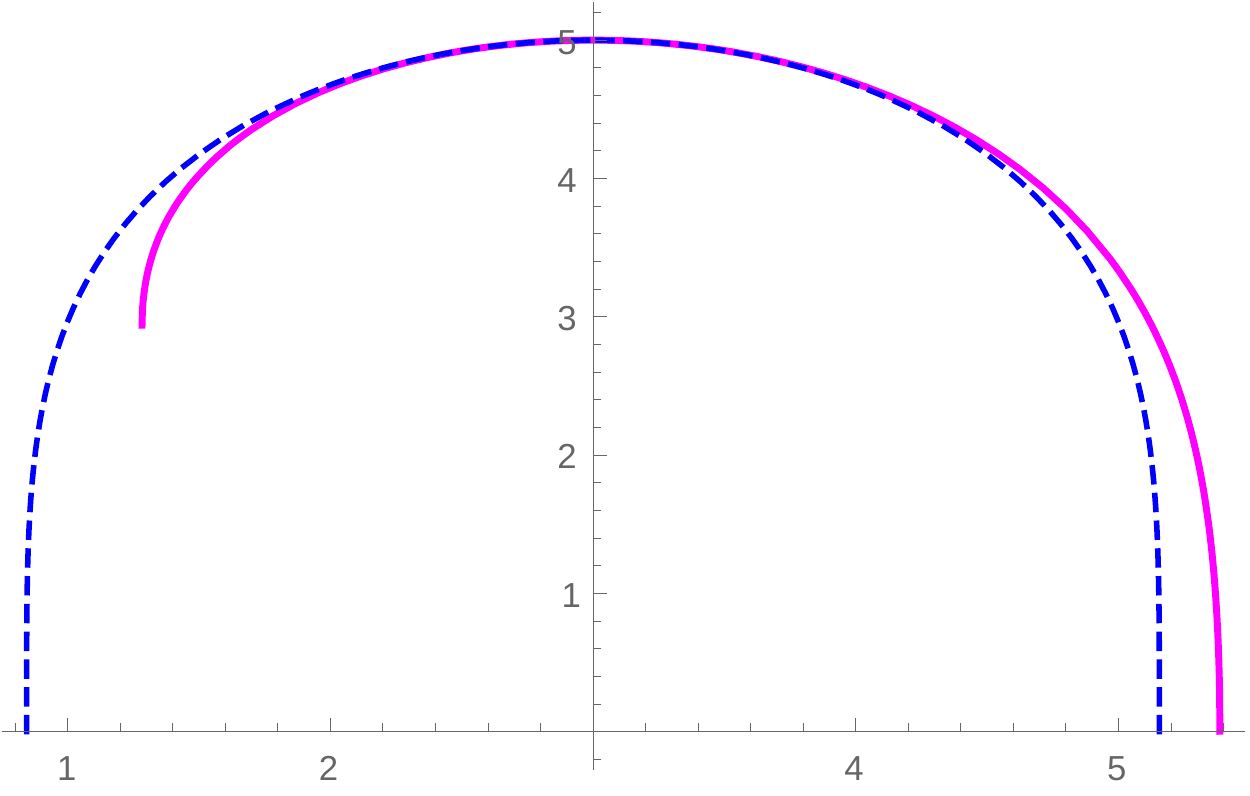}}
 
 \caption{In these three plots we take, $z_*=5$ and (a) $r_0=12$, (b) $r_0=7$, (c) $r_0=5$, (d) $r_0=3$.}
 \label{fig:2}
\end{figure}

To evaluate the entanglement entropy we need to compute the area of the extremal surface (\ref{AF}). The integral in (\ref{AF}) however yields a divergent result
which needs to be regularized. One may contemplate a suitable scheme
of regularization in the following way.
Let us consider the case of pure AdS where one can set a cut off {\textit{viz}}. $z=\delta$ near the boundary. 
The area of the extremal surface is consequently found to be
\begin{equation}
 \label{adsA}
 A = \frac{\sqrt{1-\frac{\epsilon^6}{z_*^6}}}{\epsilon ^2}-\frac{\sqrt{\pi } \Gamma \left(\frac{5}{3}\right)}{4 z_*^2 \Gamma \left(\frac{7}{6}\right)}
 +\frac{\epsilon ^4 \, _2F_1\left(\frac{1}{2},\frac{2}{3};\frac{5}{3};\frac{\delta ^6}{z_*^6}\right)}{4 z_*^6} \;.
\end{equation}
In the limit $\epsilon \to 0$, the last term in (\ref{adsA})goes to zero, whereas the first term diverges as $\frac{1}{\epsilon^2}$.  
In the same way one should be able to compute the regularized area of the extremal surface for space dependent $AdS$-Kasner geometry by subtracting 
the contribution $\sim \frac{1}{\epsilon^2}$ of the pure AdS part from the numerically calculated area.
However the computations turn out to be very hard near the singularity, even on the numerical front, although it is possible to carry out the same calculations 
in a region which reasonably far from the singularity.

\section{Discussion}\label{conc}
In this work, we studied the space-dependent version of Kasner geometries
within the premises of gauge/gravity duality. Our main result is the correlator calculated in section \ref{cod4}. A striking difference of the present 
computation from the case of the space-like singularity is the absence of any 
pole of the correlator near the time-like singularity. It is well known that for time-dependent Kasner geometries with negative Kasner exponents
correlation functions are plagued by divergences from a  pole at the singularity. On the other hand, for positive Kasner exponents 
the correlator smoothly goes to zero as it approaches the singularity, at least for large conformal dimensions. 
On the contrary in the timelike case, correlators are free from poles for both $p>0$ and $p<0$ cases. This is clear from equations 
(\ref{cor1}) and (\ref{cor2}) and from figures Figure-\ref{fig:cor1} and Figure-\ref{fig:cor2}. Hence we argue that in the present case,
there is no ``particle creation'' near the singularity. 
This is somewhat expected for a static space-time. On the other hand, our result shows this is true even in the case of strongly coupled theory.
One possible reason for the smoothness of the correlator is the following.
The metric (\ref{NewKADS}) near the boundary, say at a radial point $z=\epsilon$ is given by
\begin{equation}
ds_b^2=\frac{1}{\epsilon^2}\left[-r^{2p_t}dt^2+r^{2p_1}dx_1^2+r^{2p_2}dx_2^2+dr^2\right]\;.
\end{equation}
Now a coordinate transformation $r=e^{r'}$ forces the boundary metric to become
\begin{equation}
ds_b^2=\frac{e^{2r'}}{\epsilon^2}\left[-e^{2r'(p_t-1)}dt^2+e^{2r'(p_1-1)}dx_1^2+e^{2r'(p_2-1)}dx_2^2+dr'^2\right]\;.
\end{equation}
which essentially implies that the boundary field theory lives on the conformal frame shown above. Along a spatial direction,
say $x_1$, the proper boundary separation between the two points with coordinates $\pm \bar{x}_1$ is
 $L_{bdy}=2e^{r'(p_1-1)}\bar{x}_1=2r_0^{(p_1-1)}\bar{x}_1$. However, since $p_1<1$ always, therefore 
 as $r_0\rightarrow 0$ we find $L_{bdy}\rightarrow \infty$. As this proper boundary separation, in the conformal frame, 
 diverges, we can expect
 that the two-point correlation function should vanish. 
 
In \cite{Engelhardt:2015gta} the authors reasoned that, the appearance of pole
in boundary-gauge theory implies that the boundary field theory chooses non-normalizable states as a basis for the correlation functions.
In the same spirit, the absence of any pole in the correlators in the present case leads us to claim the boundary basis states to be normalizable. 
Again from the bulk geometry, we  observe that the full isometry group is broken. So, for
the boundary gauge theory the conformal group is also broken and it possibly implies that the boundary gauge theory is not in
ground state but in some excited state. The precise nature of the dual field theory for the space-dependent AdS-Kasner background is not very clear to us.   
One needs to develop a deeper understanding of the isometries of this class of geometries which may provide us with more insight into the structure of the dual
gauge theory. Computations of higher order correlation functions in this setup may also enrich us with more knowledge  of the structure of the boundary theory. 
To that end, further investigation is going on from our part.

\section*{Acknowledgments}
We thank Shamik Banerjee, Amitabh Virmani, Sudipta Mukherji, Sudipto Paul Chowdhury, Koushik Ray for illuminating discussions on various occasions. 
The work of SC 
is supported in part by the DST-Max Plank Partner Group ``Quantum Black Holes'' between IOP Bhubaneswar and AEI Golm.

\appendix
\section{Solution of Geodesic Equations}\label{geosol} 

Here we give a method to solve equations (\ref{suthree}). As already mentioned, 
these are equations convertible to a hypergeometric equation. we need to solve this equations around $r=0$ that means the argument of 
hypergeometric function $(\frac{r}{r_*})^{2p}$ should be around $\infty$.
So to do that we substitute $a(r)=r^{-q}$ right here. After solving we substitute back $q=-p$.
Our first equation now becomes
\begin{equation}
 r x''(r)-q r^{-2 q} x'(r)^3-2 q x'(r) = 0\;.
\end{equation}
Here ${}^\prime$ denotes derivative with respect to argument.
This equation can be solved. 
\begin{equation}
 \label{x-app}
  x(r) =\pm \frac{r^{1-p}}{1-p} \sqrt{\left(\frac{r}{r_*}\right)^{2p}-1} 
  \;\left[1-{}_2F_1\left\{1,\frac{p-1}{2p},\frac{2p-1}{2p},\left(\frac{r}{r_*}\right)^{-2p}\right\}\right]+c_1 \;.
\end{equation}
The integration constant $c_1$ can be found by setting $x(r_*)=0$. 
\begin{equation}
 c_1 = 0 \;.
\end{equation}

To solve the second equation of (\ref{suthree}), 
Define $K = z z'$. We can rewrite the equation as
\begin{equation}
\frac{dK}{dr} + a a^\prime {x^\prime}^2 K + a^2 {x^\prime}^2 +1 =0 \;.
\end{equation}
Again substituting $a=r^{-q}$ one finds
\begin{equation}
 K'(r)-q r^{-2 q-1} x'(r)^2 K(r)+r^{-2 q} x'(r)^2+1 = 0\;.
\end{equation}
We know solution for $x$. So substituting $x'(r)$ we find
\begin{equation}
 r \left[\left(r^{-2 q}-{r_*}^{-2 q}\right) {K}'(r)+r^{-2 q}\right]-q \,{r_*}^{-2 q} {K}(r) = 0\;.
\end{equation}
Solving above equation one finds
\begin{equation}
 K(r)=\frac{c_2}{\sqrt{r^{2 q}-{r_*}^{2 q}}}-r \, _2F_1\left(1,\frac{q+1}{2 q};1+\frac{1}{2 q};r^{2 q} {r_*}^{-2 q}\right) \;.
\end{equation}
$c_2$ is fixed by the boundary condition $\left.\frac{dz}{dx}\right|_{r=1}=0$,
\begin{equation}
 c_2=i \frac{\sqrt{\pi}r_*\Gamma(1-\frac{1}{2p})}{\Gamma(\frac{-1+p}{2p})} \;.
\end{equation}
Here we use $q=-p$.
$z$ is now given in terms of $K$ as $z^2 = 2 \int{dr K(r)}$. Doing this integration and then 
substituting $q=-p$ we have
\begin{eqnarray}
 z^2(r) &=& 2 c_2 r {r_*}^p \, _2F_1\left[\frac{1}{2},-\frac{1}{2 p},1-\frac{1}{2 p},\left(\frac{r}{r_*}\right)^{-2 p}\right] \nonumber \\
        &-& r^2 \, _3F_2\left[\left\{1,\frac{1}{2}-\frac{1}{2 p},-\frac{1}{p}\right\};\left\{1-\frac{1}{p},1-\frac{1}{2 p}\right\}
        ;\left(\frac{r}{r_*}\right)^{-2 p}\right] + c_3 \;.
\end{eqnarray}
Here the constant $c_3$ has to be fixed from the condition that at $r=r_0$ we have to reach boundary
$z=0$.

\end{document}